\documentclass[twocolumn,letterpaper,10pt]{article}
\usepackage{times}
\usepackage[dvips]{graphicx}

\usepackage{amsthm}
\usepackage{amsfonts}
\usepackage{amsmath}
\usepackage{array}
\usepackage{latexsym}

\newcommand{\R}{\mathbb R}
\newcommand{\N}{\mathbb N}
\newcommand{\eps}{\varepsilon}

\newcommand{\n}{\underline{n}}

\theoremstyle{plain}

\theoremstyle{definition}
\newtheorem{defn}{Definition}[section]
\theoremstyle{remark}

\begin{document}

\date{In Proceedings of IASTED Conference "Modelling, Simulation and Optimization ~MSO 2005~", August 29-31, 2005, Oranjestad, Aruba}

\title{Continuous Opinion Dynamics: Insights through Interactive Markov Chains}

\author{
Jan Lorenz \\
Department of Mathematics\\
University of Bremen\\
Nienburgerstr. 42, 28205 Bremen, Germany \\
email: math@janlo.de \\
}

\maketitle

\thispagestyle{empty}

\noindent {\bf\normalsize ABSTRACT}\newline {We reformulate the
agent-based opinion dynamics models of Weisbuch-Deffuant
\cite{Weisbuch2002,Deffuant2002} and Hegselmann-Krause
\cite{Hegselmann2002} as interactive Markov chains. So we switch
the scope from a finite number of $n$ agents to a finite number of
$n$ opinion classes. Thus, we will look at an infinite population
distributed to opinion classes instead of agents with real number
opinions.

The interactive Markov chains show similar dynamical behavior as
the agent-based models: stabilization and clustering. Our
framework leads to a 'discrete' bifurcation diagram for each model
which gives a good view on the driving forces and the attractive
states of the system. The analysis shows that the emergence of
minor clusters in the Weisbuch-Deffuant model and of meta-stable
states with very slow convergence to consensus in the
Hegselmann-Krause model are intrinsic to the dynamical behavior.}
\vspace{2ex}

\noindent {\bf\normalsize KEY WORDS}\newline {continuous opinion
dynamics, repeated averaging, bounded confidence}

\section{Introduction}

Consider a certain number of agents discussing a certain issue.
Each agent has an opinion about that issue and they try to find a
consensus. The opinion should be representable as a real number.
Thus, agents can compromise by averaging their opinions. We label
such dynamics as {\em continuous opinion dynamics}. 'Continuous'
refers to the type of opinions not to the time. This is in
contrast to several models of opinion dynamics which cover binary
opinions where agents have to decide 'yes' or 'no'.

Examples for continuous opinions are prices or the continuum form
left to right in politics. Clusters of agents in the continuous
opinion space may represent e.\,g.\ low-budget vs.\ luxury
shoppers or political parties.

There are several more or less complex models concerning
continuous opinion dynamics
\cite{DeGroot1974,JAP:chatterjee77,Friedkin1990,Flache2004}. The
recently most discussed models are the models of Weisbuch-Deffuant
(WD) \cite{Weisbuch2002} and Hegselmann-Krause (HK)
\cite{Hegselmann2002}. Both models use very similar heuristics. We
will present both models and focus on the dynamical differences in
computer simulation.

In the third section we will define interactive Markov Chains for
both models, which lead us to 'discrete' bifurcation diagrams that
will give us deeper insights for the agent-based models.

\section{Two Agent-based Models}

Opinion dynamics in the WD and the HK model is agent-based and
driven by repeated averaging under bounded confidence. The models
differ in their communication regime.

Agent-based means that the number of agents $n$ is the dimension
of the system. The opinion of agent $i \in \n := \{1,\dots,n\}$ at
the time $t\in\N$ is represented by $x_i(t) \in \R$. The vector
$x(t) \in \R^n$ is called the \emph{opinion profile} at time step
$t$. Both models propose a \emph{bound of confidence} $\eps \in
\R_{>0}$. Agents compromise with other agents only if the
difference between their opinions is less or equal than $\eps$. We
will define for both models the process of continuous opinion
dynamics as a sequence of opinion profiles.

\begin{defn}[Weisbuch-Deffuant Model] Given an initial profile $x(0) \in \R^n$ and a
bound of confidence $\eps\in\R_{>0}$ we define the
\emph{Weisbuch-Deffuant process of opinion dynamics} as the random
process $(x(t))_{t\in\N_0}$ that chooses in each time step
$t\in\N_0$ two random\footnote{With 'random' we mean 'random and
equally distributed in the respective space'.} agents $i,j \in \n$
which perform the action
\begin{eqnarray*}
x_i(t+1) &=& \left\{
\begin{array}{ll}
    \frac{x_i(t)+x_j(t)}{2} & \hbox{if $|x_i(t)-x_j(t)|\leq\eps$} \\
    x_i(t) & \hbox{otherwise.} \\
\end{array}
\right.\\
\end{eqnarray*}
The same for $x_j(t+1)$ with $i$ and $j$
interchanged.\footnote{This is a basic version of the model in
\cite{Weisbuch2002} with $\mu=0.5$.}
\end{defn}

The dynamics in the WD model are driven by pairwise compromising
restricted by bounded confidence.

\begin{defn}[Hegselmann-Krause Model]
Given a bound of confidence $\eps\in\R_{>0}$ we define for an
opinion profile $x\in\R^n$ the \emph{confidence matrix} $A(x)$ as
\[
 a_{ij}(x,\eps) :=
 \left\{ \begin{array}{cl}
   \frac{1}{\#I(i,x)} \quad & \textrm{if } j\in I(i,x)   \\
   0 & \textrm{otherwise,}
\end{array} \right. \\
\]
with $I(i,x) := \{j \in \n \,|\, |x_{i} - x_{j}| \leq \eps \}$.
Given a starting opinion profile $x(0) \in \R^n$, we define the
\emph{Hegselmann-Krause process of opinion dynamics} as a sequence
of opinion profiles $(x(t))_{t\in\N_0}$ recursively defined
through
\[x(t+1) = A(x(t)) x(t)\]
("$\#$" stands for the number of elements.)
\end{defn}

In the HK model each agent builds the arithmetic mean of all the
opinions which are closer than $\eps$ to his own.

\begin{figure}
\includegraphics[width=\columnwidth]{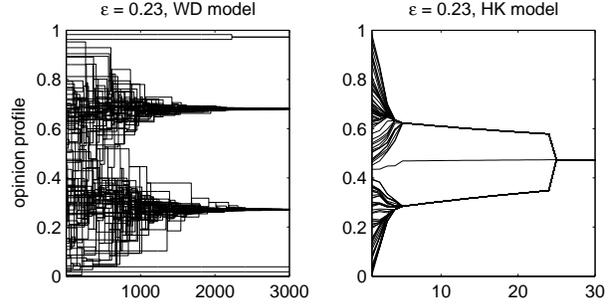}
\caption{Example processes for the WD and HK model. With initial
profile of 100 agents with random equally distributed opinions
within $[0,1]$.}\label{fig1}
\end{figure}

Figure \ref{fig1} shows example processes for both models. In both
models the processes converge to a stabilized opinion profile,
where opinions have divided into a certain number of clusters. The
stabilization can be proved analytically, even for a bigger class
of models \cite{Lorenz2005}. The dynamical behavior can be
described in the same way: The contractive force of compromising
brings opinions together, but bounded confidence forces the agents
to form clusters where higher local agent densities occur, thus
confidence to other agents gets lost.

We will briefly summarize the known results, which we consider in
this paper. At first, results about the WD model. The number of
expected major clusters is roughly the integer part of $1/2\eps$
\cite{Weisbuch2002}. But not all agents join the major clusters,
some remain outliers or form minor clusters (like in figure
\ref{fig1} at the extremes).

In the HK model there are slightly lower numbers of expected
clusters as in the WD model, for details see \cite{Urbig2004}.
Convergence to consensus may take very long, due to very few
agents remaining in the middle and attracting all other agents
very slowly (like in figure \ref{fig1} but it can be more drastic
for larger $n$).

Both models were extended in several ways. The WD model to
relative agreement, smooth bounded confidence, dynamics on small
world networks \cite{Amblard2004}, heterogeneous bounds of
confidence, dynamics of the bounds of confidence and vector
opinions \cite{Weisbuch2002}. Special starting distributions were
used to model a drift to the extremes \cite{Deffuant2002}. The HK
model has been extended to asymmetric \cite{Hegselmann2002} or
heterogeneous bounds of confidence \cite{Lorenz2003},
multidimensional opinions \cite{Lorenz2003b}. Some further
proposals are in \cite{Hegselmann2004}.

\begin{figure}
\includegraphics[width=\columnwidth]{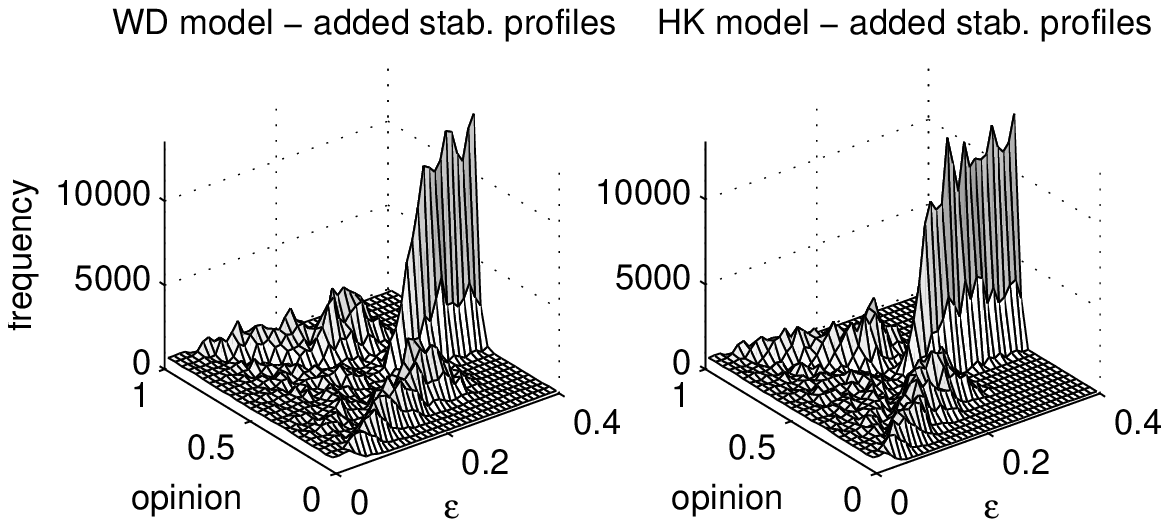} \\
\includegraphics[width=\columnwidth]{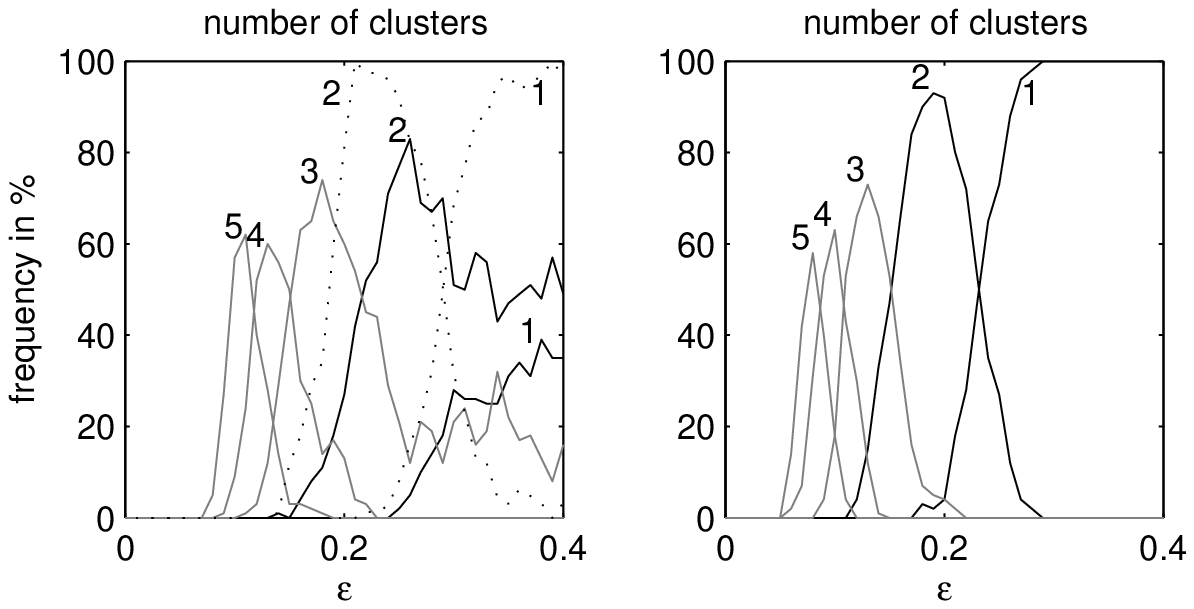}
\caption[width=\textwidth]{Added stabilized profiles (top),
frequencies of the number of evolving clusters (bottom) for the WD
(left hand) and the HK (right hand) model. Basic data: 100 initial
profiles with 200 agents and random opinions between 0 and 1;
Graphics data: stabilized profiles computed for $\eps =
0,\stackrel{+0.01}{\dots},0.4$. (Dotted lines: without minor
clusters.)}\label{fig2}
\end{figure}

We show some statistical simulation results in figure \ref{fig2}
to give a chance to look into more details. For the graphic of
added stabilized profiles we divided the opinion interval $[0,1]$
into 25 subintervals of equal size and count the number of agents
in each subinterval. In the graphic at the bottom left dotted
lines represent frequencies of 1 cluster and 2 clusters where we
count only clusters with more then 5 agents, thus outliers and
minor clusters are ignored.

The number of agents $n$ is a parameter which is not much studied
due to computing time problems. One idea of the following analysis
 studies with interactive Markov chains is to get a feeling about the
dynamics we will approach with greater $n$.

\section{Interactive Markov Chains}

In this section we want to reformulate the models of WD and HK as
interactive Markov chains. Thus, we switch from $n$ agents to an
idealized infinite population, which we divide into $n$ classes of
opinions. Instead of an opinion profile $x(t) \in \R^n$ we
consider a \emph{discrete opinion distribution} $p(t) \in S^n :=
\{p\in [0,1]^{1\times n}$ with $\sum_{i=1}^n p_i = 1\}$ as the
state of our system at time $t\in\N$. $S^n$ is the
$(n-1)$-dimensional unit simplex of row vectors. A discrete
opinion distribution is a row vector while the opinion profile is
a column vector\footnote{This is for matrix theoretical reasons:
$A(x)$ and $B(p)$ are both row-stochastic.}. For convenience we
define $p_i = 0$ for all $i\not\in \n$.

We will define \emph{transition probabilities} $b_{ij}(p) \in
[0,1]$ for agents in class $i$ to go to class $j$. For all $i\in
\n$ should hold $\sum_{j=1}^n b_{ij}(p) = 1$. Thus the
\emph{transition matrix} $B(p) = [b_{ij}(p)]$ is row-stochastic.
The \emph{interactive Markov chain} for an initial opinion
distribution $p(0)$ is the sequence of opinion distributions
$(p(t))_{t\in\N_0}$ recursivly defined through
\begin{equation}
p(t+1) = p(t)B(p(t)). \label{eq:mc}
\end{equation}
This Markov chain is called interactive (according to Conlisk
\cite{Conlisk1976}) because the transition matrix depends on the
state of the system in the actual time step.

In analogy to the bound of confidence $\eps$ we define a
\emph{discrete bound of confidence} $k \in \N$, which defines that
the transitions of opinions of class $i$ can only be influenced by
opinions of the classes ${i-k,\dots,i+k}$.

If we imagine that the opinions $1,\dots,n$ are representatives
for an equidistant partition of the interval $[0,1]$ in the way $i
\leftrightarrow [\frac{i-1}{n},\frac{i}{n}]$, we draw a heuristic
analogy $\frac{k}{n} \leftrightarrow \eps$. Thus, if we go
$k,n\to\infty$ with $\frac{k}{n} \to \eps$ we can 'converge' to an
agent-based model with $\infty$ agents, opinions in $[0,1]$ and
bound of confidence $\eps$.

In this setting we may consider $n$ as a parameter how accurate a
continuous opinion can be communicated, e.g. how many steps do we
allow on the continuous scale from minimum to maximum opinion.
(The topic of accuracy is discussed for agent-based models in
\cite{Urbig2003} in the context of opinion versus attitude
dynamics.)

Calculation of the interactive Markov chains will be interesting
regarding the following questions.
\begin{enumerate}

\item Does the Markov chain produce the same dynamical behavior as
the agent-based models?

\item What conclusions can be drawn from the results about the
idealized infinite populations in the Markov Chains to the finite
cases of agent-based systems?

\item What is the effect of the accuracy of the Markov chains $n$?

\end{enumerate}

We will start with the definiton of the WD Markov chain.

\begin{defn}[interactive WD Markov chain]
We define the \emph{WD transition matrix} for an opinion
distribution $p\in S^n$ and a discrete bound of confidence $k \in
\N$ as
\[
b_{ij}(p,k) := \left\{%
\begin{array}{ll}
    \frac{\pi^i_{2j-i-1}}{2} + \pi^i_{2j-i} + \frac{\pi^i_{2j-i+1}}{2}, & \hbox{if $i\neq j$, } \\
    q_{i}, & \hbox{if $i=j$.} \\
\end{array}%
\right.
\]
with $q_i = 1 - \sum_{j\neq i, j=1}^n b(p,k)_{ij}$
 and
\[\pi^i_l := \left\{%
\begin{array}{ll}
    p_l, & \hbox{if $|i-l|\leq k$} \\
    0, & \hbox{otherwise} \\
\end{array}%
\right.\] (Attention $i$ in $\pi^i$ is another index not an
exponent!)  Let $p(0)\in S^n$ be an initial opinion distribution.
The Markov chain (\ref{eq:mc}) with WD transition matrix is called
\emph{interactive WD Markov chain} with discrete bound of
confidence $k$.
\end{defn}

Remember that we defined $p_i = 0$ for all $i\not\in \n$. By the
founding idea of the WD model an agent with opinion $i$ moves to
the new opinion $j$ if he compromises with an agent with opinion
$i + 2(j-i) = 2j-i$. The probability to communicate with an agent
with opinion $2j-i$ is of course $p_{2j-i}$. Thus, the heuristic
of random pairwise interaction is represented. The terms
$\frac{\pi^i_{2j-i-1}}{2},\frac{\pi^i_{2j-i+1}}{2}$ stand for the
case when agents with opinion $i$ communicate with agents with
opinion $l$, but the distance $|i-l|$ is odd. In this case the
population should go with probability $\frac{1}{2}$ to one of the
two possible opinion classes
$\lfloor\frac{i+l}{2}\rfloor,\lceil\frac{i+l}{2}\rceil$.
\footnote{$\lfloor\cdot\rfloor$ ($\lceil\cdot\rceil$) is rounding
to the lower (upper) integer.}

(This definition is inspired by the rate equation in
\cite{Ben-Naim2003}. But they used continuous time. Further on we
introduced the discrete bound of confidence.)

The interactive Markov chain for the HK model looks as follows.

\begin{defn}[interactive HK Markov chain]
Let $p\in S^n$ be an opinion distribution and $k\in\N$ be a
discrete bound of confidence. The {\em $k$-local mean} of $p$ at
$i\in\n$ is
\[
M_i(p,k) := \frac{\sum\limits_{m\in\n,|i-m|\leq k} m
p_m}{\sum\limits_{m\in\n,|i-m|\leq k} p_m}.
\]
We define (with abbreviation $M_i := M_i(p,k)$ the \emph{HK
transition matrix}) as
\[
b_{ij}(p,k) := \left\{%
\begin{array}{ll}
    1 & \hbox{if $j = M_i$,} \\
    \lceil M_i\rceil - M_i & \hbox{if $j = \lfloor M_i \rfloor$, $j\neq M_i$,} \\
    M_i - \lfloor M_i\rfloor & \hbox{if $j = \lceil M_i \rceil$, $j\neq M_i$,}  \\
    0 & \hbox{otherwise.} \\
\end{array}%
\right.
\]
Let $p(0)\in S^n$ be an initial opinion distribution. The Markov
chain (\ref{eq:mc}) with HK transition matrix is called
\emph{interactive HK Markov chain} with discrete bound of
confidence $k$.
\end{defn}

Each row of the HK transition matrix $B(p,k)$ contains only one or
two adjacent positive entries. The population with opinion $i$
goes completely to the $k$-local mean opinion if this is an
integer. Otherwise they distribute to the two adjacent opinions.
The fraction which goes to the lower (upper) opinion class depends
on how close the $k$-local mean lies to it. Thus, the heuristic of
overall averaging in a local area is represented here.

For both interactive Markov chains we define a stabilized
distribution $p^\ast \in S^n$ as a fixed point (that means that it
holds $p^\ast = p^\ast B(p^\ast)$). Both models converge to
stabilized distributions for every initial distribution, while the
set of possible stabilized distributions is huge. This properties
are stated by observation of example Markov processes. There is no
formal proof at this time.

In a first analysis step we will focus on the initial distribution
$p(0) \in S^n$ which is equally distributed $p_1 = \dots = p_n =
\frac{1}{n}$. This coincides with the agent-based model if we
start with $x\in [0,1]^n$ where the $x_i$ are chosen at random and
equally distributed.

\begin{figure}
\includegraphics[width=\columnwidth]{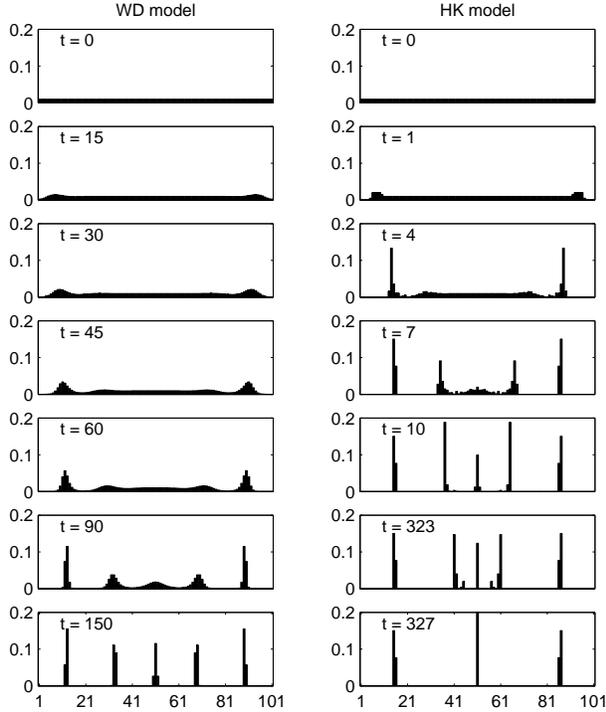} \\
\caption{WD and HK process for equally distributed initial
distribution with $n=101, k=9$ at characteristic time steps.
}\label{fig3}
\end{figure}

Figure \ref{fig3} shows processes for both models with $n=101,
k=9$ (coincides with $\eps\approx 0.089$) at characteristic time
steps\footnote{The programming of the HK model is numerically
vulnerable. Programming it like it is may lead for equally
distributed initial distributions to asymmetric distributions.
This is theoretically impossible for symmetric initial
distributions. This problem is circumvented in this calculation,
by making the distribution symmetric with
$p=(p+\mathrm{flip}(p))/2$ after each time step.
($\mathrm{flip}(p)_i := p_{n+1-i}$ for all $i\in\n$) }. We see
clustering dynamics as in the agent-based models. Five major
clusters emerging in the WD model and three major clusters
emerging in the HK model.

\begin{figure}
\centering
\includegraphics[width=\columnwidth]{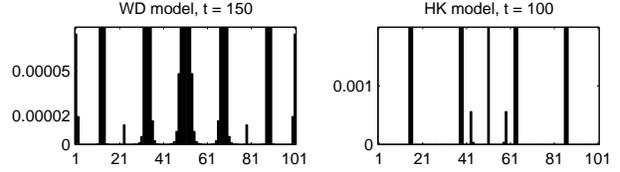} \\
\caption{Same processes as in figure \ref{fig3} with finer
$y$-scale at certain time steps to show minor clusters after
stabilization for WD and minor clusters attracting major clusters
for HK.}\label{fig31}
\end{figure}

The finer scaled plots in figure \ref{fig31} give more insights.
It shows for the WD model four minor clusters in the stabilized
distribution. (The distribution at $t=150$ is not totally
stabilized, but very close to the distribution it converges to.
E.\,g.\ the big central cluster that we see in the blow-up figure
\ref{fig31} will converge slowly to a one bin cluster. The first
two off-central clusters will converge each to a two bin cluster
as we see it already in figure \ref{fig3}.) We got minor clusters
at both extremes and two minor clusters between major clusters,
which will survive forever. The minor clusters are only visible in
the blow-up figure. There are no minor clusters between the
central and his adjacent major clusters.

For the HK model we see the possibility of very slow convergence
in the HK model on the right side in figure \ref{fig3}, which can
be explained by figure \ref{fig31}. There we see two small
clusters which hold contact between the central cluster and the
both adjacent major clusters. The dynamic reaches a stabilized
distribution with three major clusters at $t=327$. We call the
states from $t=10$ to $t=326$ 'meta-stable' because they look like
stable.

Thus, both special features that we mentioned for the agent-based
models (minor clusters for WD model and very slow convergence in
HK model) also occur in the aggregated models of the interactive
Markov chains. This effects may be seen as artifacts by simulators
but they should be regarded as intrinsic to the dynamic's behavior
(at least for great numbers of agents).

To get a complete overview about the random equally distributed
initial distribution we will calculate a 'discrete' bifurcation
diagram for both models. We will calculate the stabilized
distributions for all processes for $n=1001$ and
$k=50,\stackrel{+1}{\dots},300$ (thus $\eps\approx
0.05,\dots,0.3$). Our bifurcation diagram is called 'discrete'
because the analyzed bifurcation parameter $k$ is discrete, which
does not fit into the usual terms of bifurcation theory. In our
setting a bifurcation is an abrupt change in the stabilized
profile in one discrete step of the parameter $k$.

\begin{figure}
\includegraphics[width=\columnwidth]{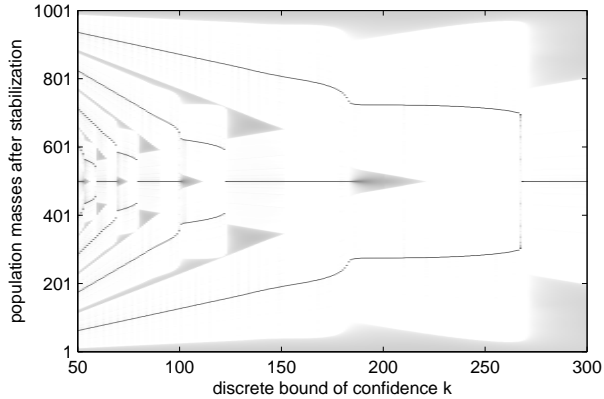} \\
\caption{Bifurcation diagram for the WD model. $n=1001,
k=50,\dots,300$.}\label{bifWD}
\end{figure}

\begin{figure}
\includegraphics[width=\columnwidth]{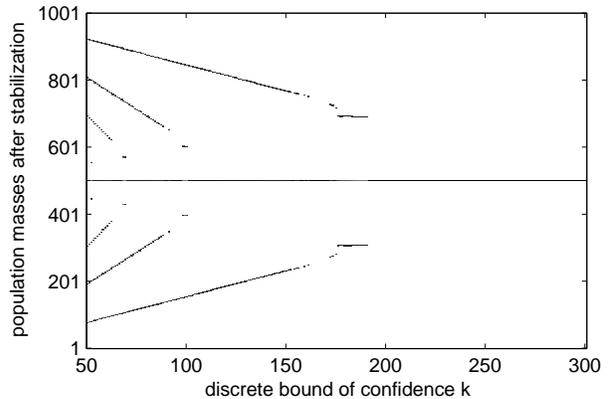} \\
\caption{Bifurcation diagram for the HK model. $n=1001,
k=50,\dots,300$.}\label{bifHK}
\end{figure}

Figure \ref{bifWD} and \ref{bifHK} show discrete bifurcation
diagrams for the interactive WD and HK Markov chains. The gray
scale symbolizes the masses at the specific positions: White is
zeros, black is fair above zero and grey is slightly above zero.
We see the major clusters as black lines. (Notice that a fully
converged cluster normally consists of two adjacent non-zero
entries in $p$ surrounded by zero entries. You can not see this in
the figure.)

We describe the bifurcation diagram for the WD process. At first
we have to notice again that the diagram is not fully converged
due to computation time. Going further on in time the gray areas
would converge slowly to gray lines of minor clusters. Going form
$k=300$ down to 50 we see at first one central cluster and minor
clusters at the extremes (that will converge to the total
extremes), then the central cluster bifurcates into two major
clusters at roughly $k \approx 270$, then a small central cluster
nucleates ($k \approx 220$), the new central cluster is growing
and the two former clusters drift away from the center. At $k
\approx 150$ two minor clusters emerge. At roughly 125 the central
cluster bifurcates again into two clusters, while the minor
clusters vanish for a short phase. The same bifurcation and
nucleation procedure repeats on a shorter $k$-scale from 125 to 80
and so on.

The existence or absence of minor clusters between major clusters
can be explained with the accuracy $n$. Only if the accuracy
leaves enough classes between two major clusters a minor cluster
can survive in between.

A similar bifurcation diagram for the differential equation
describing the same heuristics is in \cite{Ben-Naim2003}.

In figure \ref{bifHK} we see the bifurcation diagram for the HK
model. In contrast to the WD diagram the processes in this diagram
have converged completely. Going form $k=300$ down to 50 we see
one central cluster, then at $k=190$ a bifurcation into two major
clusters occurs while very little mass remains in a central
cluster. Then (k=175) we get into an interesting phase where
consensus strikes back, but not for all $k$, sometimes we end up
in three clusters. The outer clusters begin to drift outwards away
from their old position. Consensus in this $k$-phase happens due
to very slow convergence to the center like in figure \ref{fig3}.
The last consensus appears at $k=153$ (where convergence lasts
21424 time steps). Then the existence of the two outer clusters
and their outward drift is stable until the central cluster
bifurcates again on a shorter time scale (two major clusters
$k=100,99,98$, consensus strikes back $k=97,\dots,89$, then stable
5 clusters) and so on. Remarkably is that we reach a consensus at
the HK Markov chain for $\eps = k/n > 0.19$ but in the tested
agent-based model (figure \ref{fig2}) with 200 agents the chance
for consensus is only for $\eps \geq 0.23$ fair.

\begin{figure}
\includegraphics[width=\columnwidth]{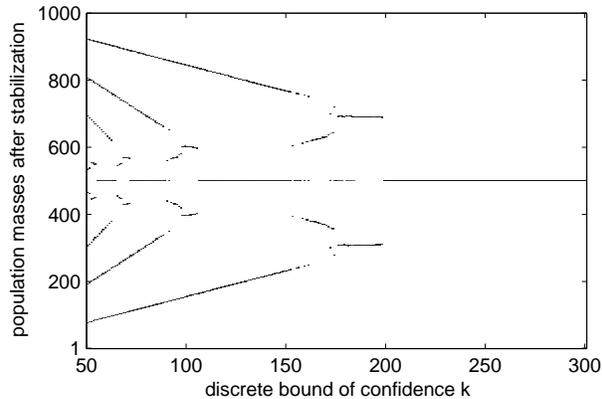} \\
\caption{Bifurcation diagram for the HK model. $n=1000,
k=50,\dots,300$.}\label{bifHKeven}
\end{figure}

The interactive Markov chains raises the question: Is it important
to have an even or odd number of opinion classes? In other words,
is it important to have a central opinion class? The impact of the
existence of a central opinion class is very high for the HK model
as we see in figure \ref{bifHKeven}. For the WD model the
bifurcation diagram for $n=1000$ looks just as figure \ref{bifWD}.

There are two differences between the HK bifurcation diagrams in
figures \ref{bifHK} and \ref{bifHKeven} which are generic for
other other pairs $n,n+1$. First, the bifurcation from consensus
to polarization occurs earlier for even $n$ (at $k=200$) and the
central clusters disappears totally at that bifurcation and
nucleates again some $k$-steps later. Second the 'consensus
strikes back' $k$-phase changes into a polarization phase, but
with major clusters closer to the center. As for $n=1001$ this is
not stable in that $k$-phase, for some values we reach two outer
majors and a central cluster.

Both phenomena can be explained through the meta-stable state that
we reach after the first time steps (in figure \ref{fig3} at
$t=10$). For an odd number of opinion classes we have a central
cluster consisting of only one opinion class, while for an even
number of opinion classes the central cluster contains two opinion
classes.

\section{Conclusion}

We summarize. The basic dynamical feature of both models is:
Stabilization to a fixed point and fixed points are opinion
distributions where the mass clusters in pairs of adjacent classes
which have distances greater than $k$ to all other pairs. This
coincides with the agent-based models (neglecting the clustering
in pairs of opinion classes). The differences in the agent based
models (number of clusters at specific $\eps$, existence of
outliers) appear also and even more drastic in the Markov chain
models. We show that two properties which may be seen as artifacts
are intrinsic to the dynamical behavior.

The WD model leads for equally distributed initial distributions
to major clusters which lie as far from each other, that it is
possible for small cluster to survive between them, but only for
great $n$ and not too low $k$. The existence of minor clusters
depends also on $\frac{k}{n}$.

The HK model leads to major clusters only (except the central
cluster, which may be small). In the center we can reach a
meta-stable state with two major clusters, a central cluster and
between major and central cluster small clusters, which attract
the big ones very slowly to the center. An even number of opinions
lowers the chances for central consensus because the central mass
can split.

In future research we should ask how robust the results are with
other initial distributions. A first but not systematic
calculation leads to the hypothesis that the basic features
(clustering in pairs, space for minor clusters for WD and meta
stable states with long convergence for HK) also characterize the
dynamics for other initial distributions. Further on stabilization
should be proved.

The heuristics of these opinion dynamic models is not based on
quantitative data. Thus, they can not predict quantitative opinion
clustering. Conclusions about real opinion dynamics should be
drawn in a qualitative manner that links the heuristics of the
model to the dynamical outcome, in a way like: Bounded confidence
in opinion dynamics about 'continuous' topics leads to clustering.
Pairwise communication produces minorities while averaging over
all acceptable opinions may lead to meta-stable situations where
consensus is possible but reaching it will take very long.

\paragraph{Acknowledgment} The author thanks the
Friedrich-Ebert-Stiftung, Bonn, Germany for financial funding.

%\bibliographystyle{unsrt}
%\bibliography{../../../lit}

\end{document}